\documentstyle[preprint,osa,aps,prl,epsfig]{revtex}
\begin{document}


\title{Mott transition observed by micro-Raman scattering in VO$_2$\thanks{Meeting Abstracts of the Physical Society of Japan 61 (2006) 575
(Presentation No:28aUE-6), (Presentation Date: March 28,
2006)({Corresponding author: H. T. Kim, e-mail:
htkim@etri.re.kr})}}
\author{Hyun-Tak Kim, Byung-Gyu Chae, Bong-Jun Kim, Yong-Wook Lee, Sun-Jin Yun, \\ Kwang-Yong Kang}
\address{IT Convergence Component Lab, ETRI, Daejeon 305-350,
Korea}
\maketitle

\begin{abstract}
A strongly correlated Mott first-order metal-insulator transition
(MIT) (or Jump) not accompanied by the structural phase transition
(SPT) was clearly revealed in VO$_2$, (New J. Phys. 6 (1004) 52,
Appl. Phys. Lett. 86 (2005) 242101, Physica B 369 (2005) 76). In
order to re-confirm the MIT for a VO$_2$-based device with a
narrow width of 3 $\mu$m and a length of 20 $\mu$m such as a rod
(Fig. A), both phonon peaks (Fig. B) by a micro-Raman scattering
with a laser beam of about 5 $\mu$m and the MIT with jump in I-V
curve (Fig. C) were simultaneously measured. A device like a rod
has less inhomogeneity. The current was restricted for
measurements. The phonon peaks of monoclinic exist even after the
abrupt jump, and disappear in over 10 mA. The jump was changed to
negative differential resistance type during Raman measurement
after the jump. The high current causes a Joule heat which arises
from the SPT near 68$^{\circ}$C from monoclinic to tetragonal. The
clean film surface without a breakdown damage after several
measurements was taken by a micro-photograph camera (Fig. A). The
MIT (jump) occurs prior to the SPT and not affected by the SPT as
evidence of electron-phonon interaction. Thus VO$_2$ is a Mott
insulator not Peierls insulator.

\begin{figure}
\vspace{1.0cm}
\centerline{\epsfysize=18.0cm\epsfxsize=16cm\epsfbox{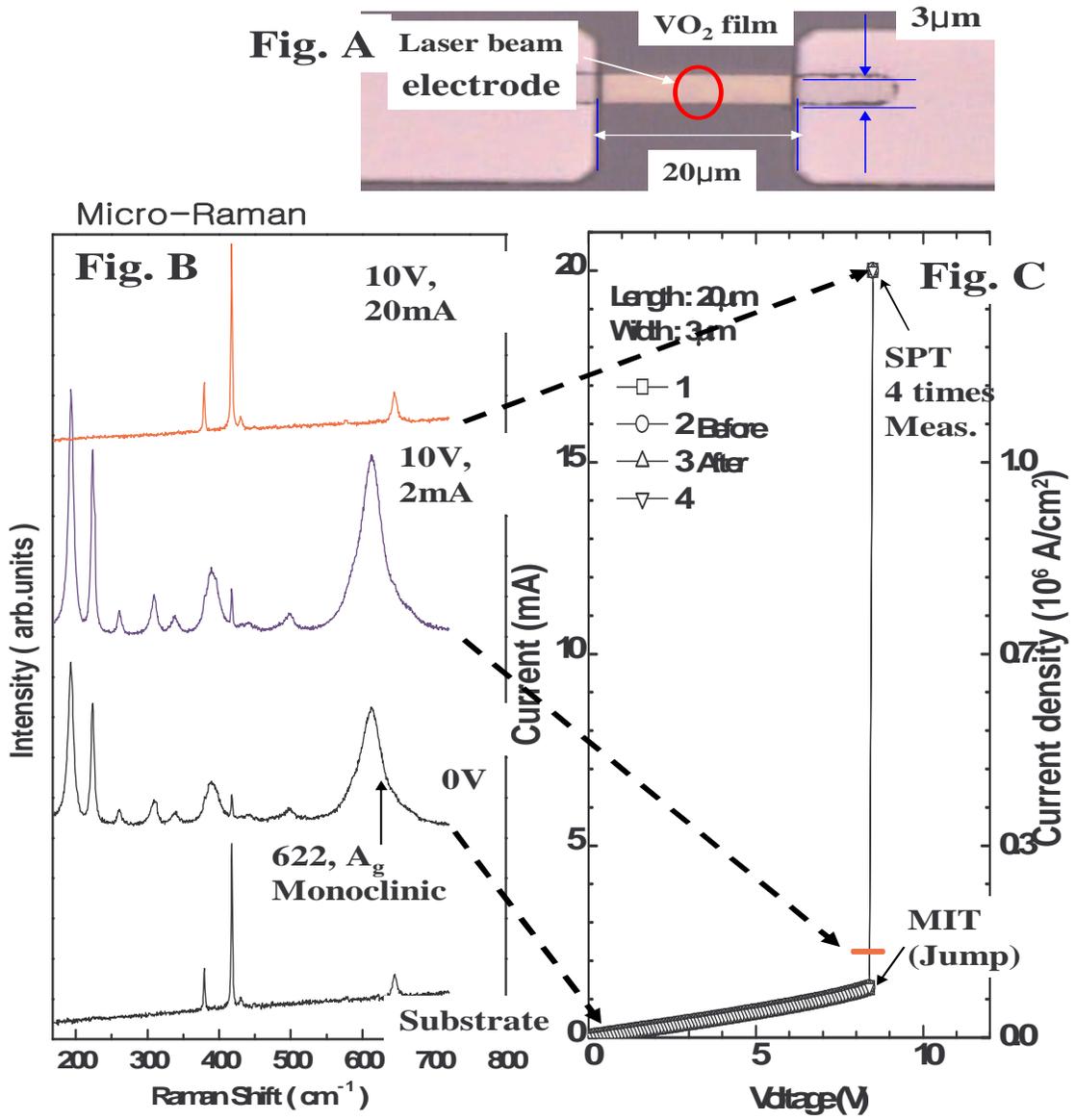}}
\vspace{0.5cm} \caption{Fig. A.: Device photograph taken after
measurements. Fig B: Micro Raman spectra. Fig. C: Jumps of two
times before and after Raman measurement were measured.}
\end{figure}

\end{abstract}

\end{document}